Democratizing Children's Computation: Learning Computational Science as Aesthetic Experience

Amy Voss Farris

Vanderbilt University, USA

Pratim Sengupta

University of Calgary, Canada

Author Note

Correspondence concerning this article should be addressed to Pratim Sengupta, University of Calgary – Werklund School of Education; Department of Learning Sciences; EDT 836, 2500 University Dr. NW, T2N 1N4 Calgary, Alberta, T2M.
Email: pratim.sengupta@uclagary.ca.

Acknowledgments

Financial support from the National Science Foundation (NSF CAREER OCI # 1150230) is gratefully acknowledged. The authors gratefully acknowledge Uri Wilensky's work on redefinition of the "concrete" as the primary inspiration behind this paper. Thanks to Anustup Basu for invaluable discussions about Heidegger and phenomenology, and Barbara Stengel, Leonard Waks and Chris Higgins for their comments and support.

Abstract

In this paper, we argue that a democratic approach to children's computing education in a science class must focus on the *aesthetics* of children's experience. In *Democracy and Education*, Dewey links 'democracy' with a distinctive understanding of 'experience'. For Dewey, the value of educational experiences lies in "the unity or integrity of experience" (*DE,* 248). In *Art as Experience,* Dewey presents *aesthetic* experience as the *fundamental* form of human experience that undergirds all other forms of experiences, and can also bring together multiple forms of experiences, locating this form of experience in the work of artists. Particularly relevant to our current concern (computational literacy), Dewey calls the process through which a person transforms a material into an expressive medium an aesthetic experience (*AE*, 68-69). We argue here that the kind of experience that is appropriate for a democratic education in the context of children's computational science is essentially *aesthetic* in nature. Given that aesthetics has received relatively little attention in STEM education research, our purpose here is to highlight the power of Deweyan aesthetic experience in making computational thinking available to and attractive to all children, including those who are disinterested in computing, and especially those who are likely to be discounted by virtue of location, gender or race.

Democratizing Children's Computation: Learning Computational Science as Aesthetic Experience

**Introduction**

Over the past several years, *computational literacy*[1] has become an important topic for discussion for K12 STEM education. Developing computational literacy requires developing epistemic and representational practices such as thinking algorithmically, and designing and creating computational artifacts such as programs and simulations. Still, computational literacy does not yet have any noticeable representation in the standard scope and sequence of public schools, especially at the elementary level. Several scholars have argued that increasing access to computational literacy for children in the realm of public education involves integrating computation with existing courses such as science and math that *all* children are required to take, rather than trying to create room for computer science as a new curricular domain[2]. Some scholars have also argued that broadening access to computation must involve efforts to create computing in the image of children's lives, and *not* vice versa; however such anthropological approaches to children's computing remain largely outside the purview of STEM classrooms[3].

In this paper, we argue — in part by example — that an effective and democratic approach to children's computing education within a science class can and must focus on the *aesthetics* of children's experience. In *Democracy and Education*, Dewey links 'democracy' with an even more distinctive understanding of 'experience'[4]. For Dewey, the value of educational experiences lies in "the unity or integrity of experience" (*DE,* 248). In *Art as Experience,* Dewey presents *aesthetic* experience as the *fundamental* form of human experience that undergirds all other forms of experience that can also bring together multiple forms of experiences, and locates this form of experience in the work of artists. Particularly relevant to our current concern (computational literacy), Dewey calls the process through which a person transforms a material into an expressive medium an aesthetic experience (*AE*, 68-69). In this paper, similar to Higgins[5], we maintain that the kind of experience that is appropriate for a democratic education is essentially *aesthetic* in nature. We extend Higgins' argument by claiming that aesthetic education should not only be exemplified by the arts – it must also bring computing and science education into its fold. Further, we illustrate that in doing so it can fundamentally transform computational science as an *experience* to a more inclusive one, especially for young learners at the fringes of computing.

This paper is structured as follows: First, we explain that a) although aesthetics has been studied as an important aspect of the work of scientists, that framing of aesthetics does not concern itself with a democratic nature of learning; and b) the discourse about children's computing and science education in general fails to account for the aesthetic dimensions of learning. Next, we articulate our own view that *democratizing* children's computing (in a Deweyan sense, that is, grounded in aesthetic experience) is a pathway to worthwhile STEM education. Finally, we demonstrate that this democratized and aesthetic experience is possible by tapping our own research as a working model, that is, a demonstration of the integration of computation as an example of how it became a democratizing force.

## On Aesthetics in Professional Science and K12 STEM

Philosophers and historians of science agree that there is an epistemic role of beauty and aesthetics in the development of scientific knowledge, and furthermore, that both beauty and aesthetics often represent deep conceptual understanding in science. For example, Kosso argued that physicists' own admissions of what they find beautiful about their theories are premised on the aesthetic qualities of coherence and interconnection, *e.g.,* a deep understanding of the relevant theory, as that may reveal "interconnectedness of facts"[6]. Along similar lines, Engler[7] argued that the beauty associated with Einstein's theories, which was also widely acknowledged by his peers, can be understood in light of the following aesthetic qualities: simplicity, symmetry (including invariance, equivalence, and covariance), unification (unity) and fundamentality. It has also been argued that choosing theories on aesthetic grounds – i.e., what makes a theory *beautiful*—is neither irrational nor a hindrance to progress because the aesthetic properties of theories are, by and large, reliable indicators for the empirical adequacy of theories,[8] and even in cases where aesthetic qualities are derived from finding analogical relationships between a multitude of phenomena by conducting "mimetic" experiments,[9] the underlying theme of interconnectedness is still evident.

Philosophers of science have also argued for the importance of scientists' interpretive work in the production scientific knowledge, including scientific inscriptions, such as drawings, diagrams, photographic images, and computer visualizations. For example, Nersessian showed the importance of developing fictive representations as explanatory models in Maxwell's work on electricity. More generally, Gooding noted that scientists make knowledge by "relocating it, moving it from the personal and local context to the larger domain of publicly reproducible phenomena, proofs, or processes."[10] In their critique of objectivity in the sciences, Daston and Galison[11] argue that the production of inscriptions in science reflect the values and the epistemology of the scientific culture in which they are made. For example, with the rise of photographic technologies in the 19th century, the perceived objectivity of mechanized images led to treating them as "facts"; the use of interpretive representations such as hand-made drawings also declined. This form of mechanical objectivity can be contrasted with the use of images in modern astronomy, where non-imagistic and necessarily interpretive representations

(for example, infrared emission data) are often represented alongside and within photographic images, in a manner that is meaningful to a wider audience[12]. The progressive centrality of computation in current scientific practice has further transformed the epistemological nature of science. For example, scientists computationally develop simulations in cases where data is sparse; these simulations, which are essentially fictive and analogous representations of reality, then serve as further sources of data.[13]

One can therefore conclude that although scientists generally acknowledge that their work has inescapable aesthetic dimensions, they have typically focused on the aesthetics of "final form" science. There is little, if any, understanding of the aesthetic dimensions of scientific practice, as well as the journey of "becoming" a scientist. In the domain of K12 STEM education research, a few scholars have begun recognizing that learning science is itself an aesthetic experience; however, this body of work, despite adopting a Deweyan perspective, proposes a rather thin, and purely discursive conception of both aesthetics and experience.

For example, Girod, Rau and Schepige[14] adopt the transformative and continuous qualities of Deweyan aesthetic experiences. They argue for including the artful aspects of science for generating interest and an expansion of perception in the sciences. They locate aesthetics in the classroom discourse, and present guidelines for teaching science as an aesthetic experience by proposing forms of questions, that teachers could ask students during the curriculum, with the central goal of helping students establish personally meaningful connections to content. Jakobson and Wickman[15] also adopted a similar definition of aesthetics and identified the role of qualitative judgments such as "nice", "disgusting", etc. in students' utterances as indicators of conceptual understanding, and genres of experiences.

While applauding the inclusion of aesthetic experiences in the K12 science classroom,

Lemke critiques this approach by arguing that "the heightened vitality" we associate with a Deweyan "Experience" in professional science can be understood in terms of "accounts of what happened, or experienced from idea to design to data and conclusions"[16]. Lemke argues that central to this account of the transformative nature of Experience in the production of scientific knowledge is "the vital fusion of theory and experiment (or observation) that makes science truly a performance art," which science education has failed to address in an authentic fashion.

We agree with Lemke's critique, that in science education research, the transformative nature of experience resulting from fusion of theory and experiment is rarely investigated. The historian of science Andrew Pickering has termed this vital fusion the "mangle of practice," and argued that this fusion is much deeper than discourse, by showing that at the heart of scientific progress is the "dance of agency" between theories and instrumentation.[17] However, Pickering's work does not address the dimensions of the scientists' affective involvement and personal meaningfulness, which are also essential elements of Deweyan aesthetic experiences. Along this line, a further critique of the studies of aesthetics in professional science is that they can be viewed as efforts to identify "beauty at the helm", as they focus on the aesthetics of the interested—*i.e.*, accomplished scientists, who were deeply interested and thoroughly engaged in their professional pursuit. A heightened form of deep engagement occurs when the scientist literally identifies herself or himself with the object of inquiry by conceptually projecting herself or himself on the object, and engages in thinking *like the object of inquiry.*[18]

In stark contrast to such heightened forms of engagement and experiences lie the learning experiences of the dis-empowered and the dis-interested, who are typically left out of the fold of deep engagement with the curricular content in most classrooms.[19] In the context of educational computing, this population includes women and ethnic minority students as well as students interested in the arts, most of whom do not identify themselves as computing or STEM competent, even at the college level.[20] As Dewey argues, democratizing education necessitates the focus on how to foster conditions where such students will develop a deep interest in their curricular work, and (in the context of learning science) come to see that work as both scientifically and personally meaningful (*DE*, for example, 128, 227-239). We therefore posit that the study of aesthetic experience in science learning that does not concern the disinterested or the disenfranchised is fundamentally undemocratic. We therefore ask the following question: What is the nature and role of aesthetic experiences for such students in the context of doing computational science? We address this in the next section.

**Democratizing Science and Computing Education: The Nature and Role of Aesthetic Experiences**

Dewey argued that since a democratic society repudiates the principle of external authority, it must find a substitute (of authority) in voluntary disposition and interest, and further, that education is the means through which interest could be generated (*DE*, Chapter 7). Achieving coherence between the learners' interest or what he terms "true aims", and pedagogical aims, would foster continuity of the pedagogical experience with the learners' experiences outside the classroom (*DE*, Chapter 8). Dewey therefore argues for two forms of continuities—continuity of the curricular experience with the learner's life outside the classroom, and continuity of the pedagogical aims with the "true aim" of the learner. He claims that the latter form of continuity is dependent on the former.

For Dewey, "the measure of the value of an experience lies in the perception of relationships or continuities to which it leads up" (*DE*, 147). The richness of an experience is marked by a variety of interests, but Dewey argues that these interests have been "torn asunder" in schools. Curricular domains of knowledge are institutions that are disconnected from each other (*DE*, 294-297), and this isolation of curricular experiences "rupture[s]… the intimate association" between domains of knowledge as experienced by the learner in a continuous form in his or her everyday life outside the classroom (*DE*, 295). Dewey considers this a serious breach in the learners' continuity of mental development, because this makes the curricular experience unreal for the learner, and can therefore, lead to a loss of interest. Dewey then challenges us to think beyond these discontinuities for pedagogical design:

> The point at issue in a theory of educational value is then the unity or integrity of experience. How shall it be full and varied without losing unity of spirit? […] How shall we secure breadth of outlook without sacrificing efficiency of execution? How shall we secure the diversity of interests, without paying the price of isolation? (*DE*, 238-239)

In contrast to the fragmented experiences that are still common in public educational settings stands a more fundamental form of experience that in his later work, Dewey termed "[a]esthetic experiences."[21] Dewey argues that in the case of an aesthetic experience, the traditional divide between domains of knowledge (such as science, art, religion, etc.) do not exist, because such experience is fundamental to *all* domains. He finds the paradigm of such experiences in the artist, and argues that aesthetic experiences arise in the artist's process of transformation of a *material* into an *expressive medium* (*AE*, 111-113). The process of expression is necessarily constrained, but not restrained—that is, the conversion of an act of immediate discharge (i.e., a direct representation) into one of expression depends upon the existence of conditions that impede direct manifestation and instead "switch it to a channel where it is coordinated with other impulsions" (*AE*, 102). This modification of the original impulsion by "cooperative" and "collateral tendencies" gives it added meaning – "the meaning of the whole of which it is henceforth a constituent part" (*AE*, 102). The expressiveness of the object therefore represents an interpenetration of the materials of undergoing and of action, and thus, the "complete fusion of what we undergo during the process of expression" (*AE*, 108).

It is this interpenetrative nature that makes aesthetic experiences *fundamental,* in that they transcend domains of knowledge and represent the unity of experience through which the object becomes expressive, and personally meaningful to the artist. Aesthetic experiences thus foreground *experience* over canonical forms of knowledge that typically exist in isolation from one another, both in professional practice and pedagogy. This isolation, Dewey argued, is the result of "non-experiential" or "anti-experiential" philosophies,[22] which Dewey contrasts with the fundamentally continuous nature of experience.

We find Dewey's notion of *aesthetic experiences* to be appropriate for our purposes for two reasons. The first reason is tied to the nature of computation (including its practice): domain-generality is a "habitual nature" (*AE*, 109) of computational programming and modeling. The creation of computational programs that underlie any usable software (or application) involves the use of computational abstractions,[23] *i.e.,* representational structures that are domain-general (e.g., algorithms, data structures such as lists and arrays, etc.). That is, the same programming language can be used to create applications in diverse domains such as physics, biology and social sciences. In our own research, we have used the same programming language to develop models in physics, biology, microeconomics and artist networks. The essential nature of the practice of computation is therefore *transformative.* That is*,* in Dewey's terms, the material of computation—typically, a programming language—gets transformed to an expressive object, a software application that has value because of its usability and meaningfulness in other domains.

Our second argument concerns Dewey's emphasis on the continuity of learning experiences for a democratic education. Herein lies an important affordance of the particular genre of computation we use in our worked example: agent-based computation, i.e., a form of computation where a user can simulate a complex phenomenon (e.g., a traffic jam) through programming the behaviors of virtual agents, by assigning them simple, body-syntonic "rules", (e.g., moving forward, slowing down, etc). The complexity of the overall phenomenon (e.g., the formation and backward propagation of the jam) emerges from the aggregation of simple, agent-level behaviors. Furthermore, because a computational agent is a protean agent, it can take on any form: an image, a word, an object, a mathematical representation (*e.g.,* a graph), etc. This in turn makes agent-based computation a suitable medium for modeling phenomena in domains as diverse as physics, biology, art and engineering.

Over the past three decades, research on making agent-based computation accessible to young learners has identified several activity forms that can potentially support interest-driven computing. These studies extend the range of learning activities beyond the traditional image of programming as writing code to include new forms of activities within which programming is embedded: game design[24], digital narratives[25], digital animations of sketches and graphic design,[26] and integration of programming with physical computing and the use of low-tech objects.[27] Using Wilensky's[28] definition of "concrete", where concretion is defined as the process of the new knowledge "coming into relationship with itself and with prior knowledge", such forms of knowing can be termed "concrete". That is, as the learners (in these studies) engage in the development of multiple, personally meaningful representations of the object of inquiry, they begin to "see" the unknown using experiences that are personally meaningful and familiar.

One can therefore argue that these studies present us several images of learning that allude to some elements of the Deweyan notion of aesthetic experiences. For example, taken together, these studies suggest that computation, and in particular agent-based computation, is indeed a malleable medium that can lend itself to multiple activity forms, and further, that certain forms of computation might even bring together multiple domains within the act of learning. Some of these studies also show that using agent-based computation, learners can appropriate the goals of the assigned activity in order to pursue something rising from their own interests, but without losing focus on the disciplinary learning objectives. Azevedo[29] termed these forms of learner-generated activities "personal excursions."

To summarize, we argue that the *transformative* and *fundamental* nature of aesthetic experiences can provide us useful guidelines for designing an inclusive and democratic pedagogy for kids' computing in particular. Along the first dimension, we posit that pedagogical experiences should provide learners opportunities to transform a material (*e.g.,* a computational programming language) into an expressive medium. In the context of computing education, this means that the learner should be able to create a personally meaningful artifact. This in turn, requires balancing the leaners' interests or true aims with institutionally mandated aims that instructors have to abide by. With respect to the second dimension, the fundamental nature of aesthetic experiences implies that the learning experience must also be continuous. That is, it should also enable learners to connect the present experience with their lived experiences outside the classroom, and also to bridge different domains that are traditionally taught as ontologically distinct from one another. The example offered here represents a computing experience that is both transformative and fundamental for Matt and Ariana. The experience was inclusive, inviting them in to a domain of practice for which they initially had no interest, and enabled them to participate fully.

**A Worked Example**

Matt and Ariana, the two 5th grade students considered in this example were enrolled in a two week long summer course on agent-based computer modeling for learning science that we (the authors) co-taught at Vanderbilt University. During the first couple of days in the course, Ariana and Matt each disclosed to the researchers that they had no interest in computer programming. Ariana was especially interested in history and literature, and Matt was an aspiring actor. Neither saw themselves as people who might enjoy or be good at computer programming, and both of them had joined the course based on their parents' insistence.

From our perspective, as instructors of the course, the central disciplinary learning goals for students in terms of learning programming and physics were: a) to develop fluency with agent-based programming and modeling motion as a process of continuous change; and b) in the process, begin to develop deep conceptual understandings of the relationships among distance, speed, and acceleration. Developing an understanding of motion as a process of continuous change has been shown to challenging for K12 learners, particularly at the elementary and middle school level.[30] In the first phase of activities, students were introduced to the ViMAP software (described below) by drawing shapes. In Phases II and III students generated data about motion by acting as the "agent" in "real-life" situations such as travelling on the building's elevator and observing the free-fall of a block of ice. Note that our goal was to reframe learning computational science as an aesthetic experience; therefore, we intentionally integrated multiple domains and tools in our pedagogy. For example, in Phase IV, besides ViMAP, students also used a musical programming software called Impromptu TuneBlocks (described below) to build computational models of motion, based on the data they generated during the embodied modeling activities. TuneBlocks enhanced the representational palette of learners to include musical attributes such as pitch and tempo as possible representations for speed and acceleration.

For example, an object that is accelerating at a steady pace could be modeled musically in terms of the steadily increasing pitch of a note.

**The "Tools": ViMAP and TuneBlocks**

ViMAP (Figure 1) is an agent-based visual programming language and modeling platform[31] that uses NetLogo[32] as the simulation engine. Instead of typing text-based commands, users use a drag-and-drop interface to select and choose commands from a library of commands in order to control the behavior of a single computational agent—a "turtle". The ViMAP version used in this study had two components: a construction world, where learners construct their programs by organizing the visual programming blocks; and an enactment world, where a protean computational agent (or a set of agents) carries out users' commands through movement on the computer screen.

SUGGESTED LOCATION FOR FIGURE 1

Impromptu[33] is a computer-based musical programming environment in which students can learn to compose melodies using their musical intuitions by arranging small blocks that represent musical notes. We used one of Impromptu's five "PlayRooms", called Tuneblocks. In this course, students composed tunes or melodies by editing existing tunes from the TuneBlocks library and used pitch or the duration of the notes in order to represent constant speed, constant acceleration and constant deceleration (see Figure 2).

In our research, we used an illustrative case-study approach[34] grounded in naturalistic inquiry methods. We videotaped in-depth interviews with the students in order to understand students' perspectives on and explanations of their own work. These interviews were transcribed and analyzed inductively using the double-coding method in order to identify salient themes. Here we present two episodes of Matt and Ariana's work, one near the beginning of the course, and a second episode occurring during their final project. In each episode, we identify two themes, which are key criteria of Deweyan aesthetic experiences: the synthesis of multiple domains of knowledge and practice that traditionally remain separate in classroom instruction, and the balancing of true aims and institutional aims, through the realization of the representational properties of the different forms of computation media.

SUGGESTED LOCATION FOR FIGURE 2

**Episode 1: Programming "Thomas"**

After completing the introductory activity of drawing some simple Logo-based shapes, Ariana began writing a ViMAP program to make the turtle write "Thomas" (Figure 3). This activity began as a teacher-directed task, in which we asked students to either draw a shape of their choosing, or draw one letter from their name. Ariana's work spanned several days during the first week: she worked on other assigned tasks and kept returning to complete the Thomas program when she found time. Writing "Thomas" became an important side-project for Ariana, one of her own choosing.

After observing her eager work on this project, a researcher interviewed her about the meaning of an inscription of the name "Thomas." In this interview (transcript provided in the appendix), Ariana explains her relationship with Thomas, and her statements provide evidence of the continuity between her biographical experiences outside her classroom and the programming activity. There are five Thomases in Ariana's life, including the newly found ViMAP turtle.

Ariana: The fourth Thomas is my best friend, the third Thomas is from *The Maze Runner*, and the second Thomas is the Maze Runner's dad, and the first Thomas is Thomas

Edison, the scientist. And this will be our fifth Thomas, our little turtle here. And he is so good. He is going to preschool and he is knowing how to spell his name.

One way to interpret Ariana's work is that through this activity, she brings together some of her favorite aspects of her personal life—e.g., her fondness for her best friend, who is also her neighbor and a classmate and "really, really close" to her, and her favorite fiction character from a young-adult book series (Thomas in *The Maze Runner*)—and merges them with the protean ViMAP (Logo) turtle. The ViMAP turtle, as Ariana points out in her interview, is the fifth, and youngest Thomas in her life. The turtle has now become an object of affection for her - she positions Thomas the ViMAP turtle as a preschooler, who learning how to spell his name. The turtle, as Seymour Papert [35] pointed out, therefore acts as a *transitional object*—i.e., both as a protean computational object, as well as a representation of the child's favorite aspects from her own biography outside the classroom.

SUGGESTED LOCATION FOR FIGURE 3

The Thomas narrative also created a space for humor between Matt and Ariana. During her interview, Matt, another 6th grader who sat next to Ariana, jokingly complained about "too many Thomases!" As the interview began, Matt attempts, humorously, to prevent Ariana from going "through the list.". As the interview proceeded, he exclaimed "it burns! It burns," covering his ears, and making humorous expressions, playfully communicating that he did not see the importance of the Thomases that Ariana did. As Matt clarified after the interview, his attempts at humor were directed to indicate that he had already been subjected multiple times to listening to the long legacy of the Thomases in Ariana's life. This further suggests the importance of Thomas in Ariana's life. Once Ariana completed her excursion, Matt and Ariana, who did not know each other prior to this class, chose each other as programming partners and continued to work together on all subsequent assignments.

In what follows, we highlight the two key criteria of an aesthetic experience that are central to a democratic education, as evident in this episode:

*Continuity across Domains*. In Ariana's work, geometry and programming were deeply intertwined with one another. Programming involved the successful use of relevant computational abstractions, such as variables and loops. In terms of learning geometry, using turtle graphics to create the shape of a letter involves thinking *like* the turtle, in order to use the egocentric coordinate system in ViMAP, a feature of agent-based modeling. However, note that it was her love and affection for the many "Thomases" in her life that created this context for productive unification of these domains. Simulating the trajectory of the computational agent (the turtle) in the shape of each letter involved significant complexity in terms of figuring out both the turtle's egocentric coordinate system, as well as the Cartesian coordinates of the pixels at the beginning of each letter. On the other hand, the instructor-mandated activity of drawing only a single alphabet letter would have involved a far less extensive exploration of both key geometry and programming.

*Balancing Institutional Aims and True Aims*. Ariana's project shows that the computational agent (the turtle) truly became a transitional object—i.e., she projected her identity onto the turtle. Her way of learning programming was by making the turtle learn how to write Thomas. This in turn transformed the material (ViMAP) and the activity (learning programming by drawing letters) into a means to talk about her serendipitous encounters with the many Thomases in her life: literary figures, historical figures, and friends. Matt became humorously critical of Ariana's personal attachment to Thomas and her persistence with the Thomas project. This relationship created a space for playful humor between the two students, which was important for Matt, who wanted to be an aspiring actor, one with an expressed interest in comedy. As Matt became familiar with Ariana's project, he progressively developed a deep interest in how Ariana had calculated the size each letter in relationship to the geometry of the ViMAP world, because it was closely related to a challenge he was facing in his own work. During their collaboration, humor played an important role, and established a comfortable working relationship between the dyad. They decided to work together as partners for the remainder of the course.

**Episode 2: A Collaborative, Multimodal Model of Acceleration**

During the final phase of the activities, students were asked to represent how the speed of a car on a roller coaster (as shown in a YouTube video) was changing, using either ViMAP or TuneBlocks. Matt and Ariana decided to collaboratively develop both a ViMAP and a TuneBlocks model. Their ViMAP model (Figure 4) represented a period of constant acceleration of the roller coaster using line segments (dot-traces) of different colors to represent distance traveled in each interval of time. Speed was represented by gaps between successive dots: constant speed meant equal gaps between successive dots, and acceleration meant increasing gaps. Ariana also explained the significance of the color changes of the lines: "The color kind of rapidly changes and then spreads out." When one of the interviewers asked her to explain this more, she explained that:

> Ariana: …It kind of changes because it is kind of slow during HERE (pointing to top portion of the line), then it spreads out and the colored lines get further, so that would be one of the reasons that it is better [than alternative models] and it has…(pause)...This would be acceleration, see here how it is getting, how it's kind of slow, how they are all crumpled up, and they get bigger and bigger.

From the perspective of learning physics, the learning goal of the activity was for students to begin to distinguish among distance traveled, speed, and acceleration. In Ariana's explanation, "crumpled up" was a visual metaphor for slow, and "spread out" for fast. Her explanation also makes explicit how she was using a systematic change in color to represent the rate of change in motion. While she did not explicitly identify rate of change, her explanation suggests that she was beginning to identify how fast (or slow) the color was changing as an important and communicative aspect of her representation.

Later in the interview, when Amy asked Matt and Ariana about the regularity of the change in distance per unit time, Ariana pointed to the steady regularity of the placement of measurement flags in the execution of her model: the same commands repeated in a loop: forward (step-size), plant-flag, speed-up (increase of step-size), change-color (amount). The words in parenthesis indicate the parameters associated with the commands that the students also had to specify.

Matt's explanation of steady, however, was somewhat different:

Matt: Because it's just accelerating like .. [Matt snaps two of his fingers twelve times. The frequency of snaps increases steadily].

In this excerpt, Matt uses a steadily increasing frequency of a particular sound to explain what he means by "steady". The increases in distance occur regularly, because their ViMAP model is incrementing the distance travelled by the turtle in each step by the same amount that is decided using the "speed-up" command. Matt's explanation of a steady pace used a combination of gesture and sound to represent a *steadily increasing* tempo in order to represent a steady acceleration. This in turn was similar to the representation of acceleration in their TuneBlocks model that accompanied their ViMAP model. In their TuneBlocks model (Figure 2), they used two variables—pitch and duration—as representations of change in speed. They used a gradually decreasing pitch to indicate the decreasing altitude of a roller coaster moving down a steep incline, and programmed the duration of each note to represent speed, where increasingly shorter, closer together tones indicated increasing speed.

SUGGESTED LOCATION FOR FIGURE 4

After constructing both the models, Matt and Ariana decided to synchronize their models so that both the ViMAP and the TuneBlocks models would each serve as components of one unitary model of the motion phenomenon. This eventually resulted in a "live performance" (as Matt explained to an instructor), where they "played" both their models simultaneously for the instructors. During the final segment of the interview, Matt extended the description of the ViMAP model to include periods of gradual slowing down and of rest in order to illustrate a narrative about accelerating onto a highway, then getting off at an exit, stopping at a red light, and parking—a situation that was familiar to him from his daily life.

Again, we highlight the two key criteria of an aesthetic experience that are central to a democratic education, as evident in this episode:

<u>Continuity across Domains</u>. In this episode, learning about the physics of motion—i.e., learning to represent motion as a process of continuous change was deeply intertwined with use of programing and musical notations. Similar to Episode 1, the use of computational abstractions such as variables and loops continue to serve an important role here: color and gaps were used as representations of speed and acceleration. In Ariana's case, her explanation used visual attributes such as color, while Matt used a steadily increasing tempo of finger-snaps to represent rate of change. Mathematically speaking, Ariana and Matt also began to develop representations of *rate* of change of motion and represent changes in speed in terms of continuous change in the distance traveled in each successive increment of time. They each describe a different aspect of the uniformity of increase in motion that is accelerating at a constant rate: the uniformity of the chronology of measurement (Ariana) and the uniformity of the *change* in speed (Matt). Furthermore, as Ariana's verbal explanations make explicit, she was also beginning to distinguish between different rates of acceleration (e.g., "crumpled up" vs. "spread out"). The introduction of TuneBlocks further widened the representational palette for the children; musical attributes such as tempo and pitch were used as representations of speed, and modeling motion was transformed to musical composition.

*Balancing Institutional and True Aims.* The multi-modality in Ariana and Matt's models illustrate children's agency in interpreting and symbolizing scientific ideas. Contrary to the instructors' advice, the dyad also refused to choose one programming environment over another, and instead, created for themselves the goal of synchronizing two models to be executed at the same time. They therefore created a perceptually enhanced, representational account of motion that consisted of both visual and auditory representations of motion as a process of continuous change. Their final project was therefore a coordinated *performance:* Matt ran the visual simulation designed in ViMAP while Ariana played the audio tune designed in TuneBlocks in a synchronized fashion. To the students, the final project was therefore a work of art, *despite* being a composite model of motion. It was thus a realization of the true aims of the two artistically inclined children—Matt the future actor, and Ariana, the history and literature fan.

## Conclusion

Can the disinterested find their voices in the STEM classroom, especially in classrooms where computation is the medium of "doing" science? To answer this question, we have argued that one must reimagine learning computing in the science classroom as an aesthetic experience

(in the Deweyan sense). That is, the democratization of computing hinges on designing pedagogies that enable the learner, especially the disinterested, to transform the computer as a material into an expressive medium, in a manner that can create a fundamental and unifying experience for the learner. Grounded in Dewey's work in *Democracy and Education*, we have further argued that such aesthetic experiences must bring together sanctioned domain knowledge with the leaners' experience and intuitions from their everyday lives.

This bringing together is not merely an act of recognition – but as Higgins argues, it is an act of heightened perception, an act of "seeing more" rather than merely seeing. Computational media, the kinds we reported here can concretize this metaphor of "seeing more" by enriching the perceptual engagement of the learner. To this end, Matt and Ariana's work shows that by opening up the representational palette to include multiple modalities of expression such as visual dynamics of agent-based simulations, visual and auditory representations of musical notations, and musical composition, computational media can support authentic engagement of children with the analogical, interpretive, and symbol-laden work of science, while accommodating their interests. It can therefore provide children entrée into a new *kind* of science, where the mundane is reimagined and (re)represented as complex, and children's true aims find a place alongside the institutionally mandated aims.

Dewey argues that the habitual nature of art is such that it helps us "see" complexity in the world of our everyday experiences, the mundane. He wrote: "Art throws off the covers that hide the expressiveness of experienced things" (*AE*, 110). We see this re-imagining of the mundane to be at the core of democratic pedagogy, both in general, as well as in the specific context of educational computing, especially in the science classroom. The many Thomases in Ariana's life and Matt's experience of his daily car rides as well as his aspiration to be an actor are representations of children's interests and Deweyan "true aims" that are typically left behind in pedagogical time. The meaning that learners develop in STEM classrooms, in the truly democratic sense, must not be depleted or devoid of these true aims, because the "value" of a democratic education lies in the unity (or integrity) of experience. The essence of technology being nothing technological[36], we believe that the unity of experience, which Dewey argued to be at the heart of aesthetic experience, is truly the essence of democratizing children's computational science.

Appendix: Ariana and Matt on "Thomas"[37]

Ariana: Well, my best friend's Thomas and I really, really, really, we're really, really close. See, he's like my neighbor [and he's also my classmate

Matt: Don't go through the list!] Don't go through the list!

A: Okay=

M: =too many Thomases, way too many Thomases

A: And, [oh yeah, I made a video about how

M: *(whispering to the camera)* too many, too many]

A: I made a video about an [explanation of the Thomases

M: *(moaning)* I said don't go through with it!]

A: So it's Thomas *(counting on fingers)* named after Thomas who's named after Thomas, who's named after Thomas.

M: *(covering ears)* it buuuuurrrr-urrrr-urrrr-urrrrns!!

A: The fourth Thomas is my best friend.

M: It burns! It bur-ur-urns!

A: The third Thomas is..um..from *The Maze Runner,*

M: *(whispering)* It burns!

A: and the second Thomas is the Maze Runner's dad and the first Thomas is Thomas Edison..the scientist. And this will be our fifth Thomas, our little turtle here and he is so good..He is going to preschool and he is knowing how to spell his name.

Pratim: So the turtle is going to preschool and he is knowing how to [spell his name?

M: Heeelp me, help me! (hands extended to the camera)]

A: Yes, basically.

Pratim: Alright.

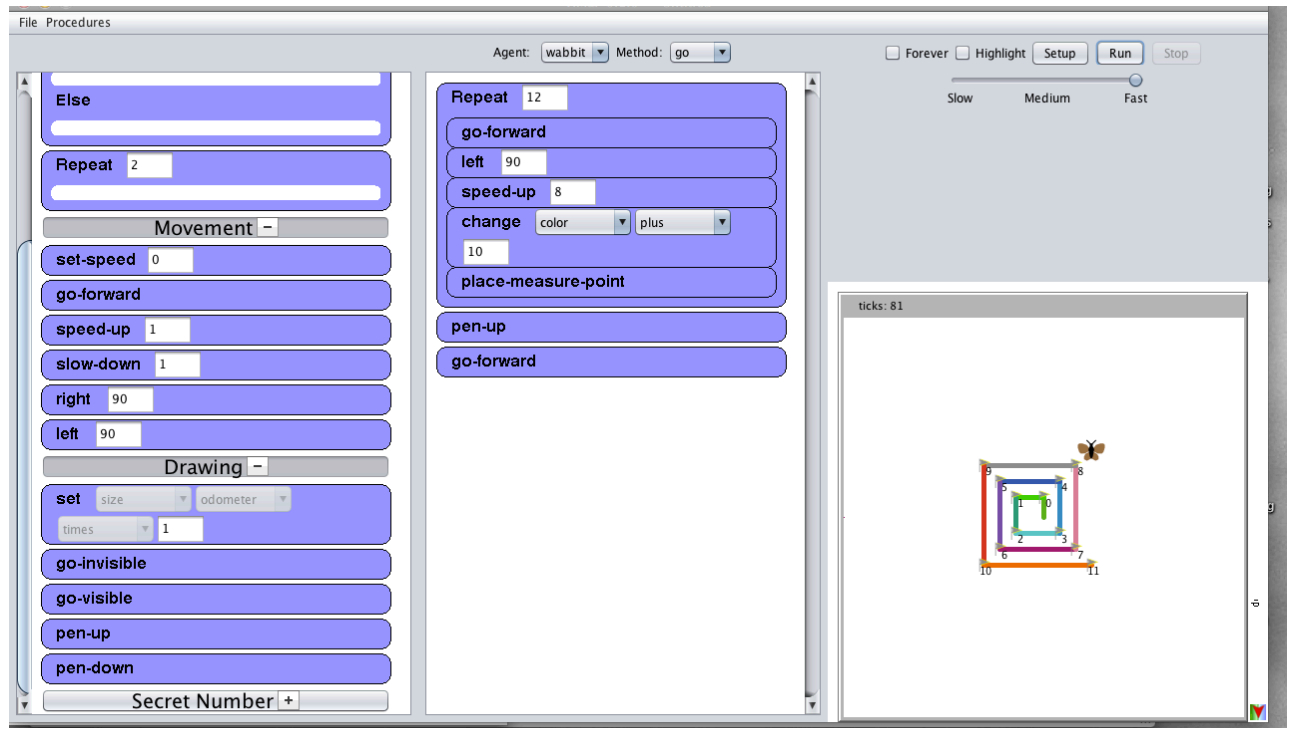

*Figure 1: The ViMAP interface*

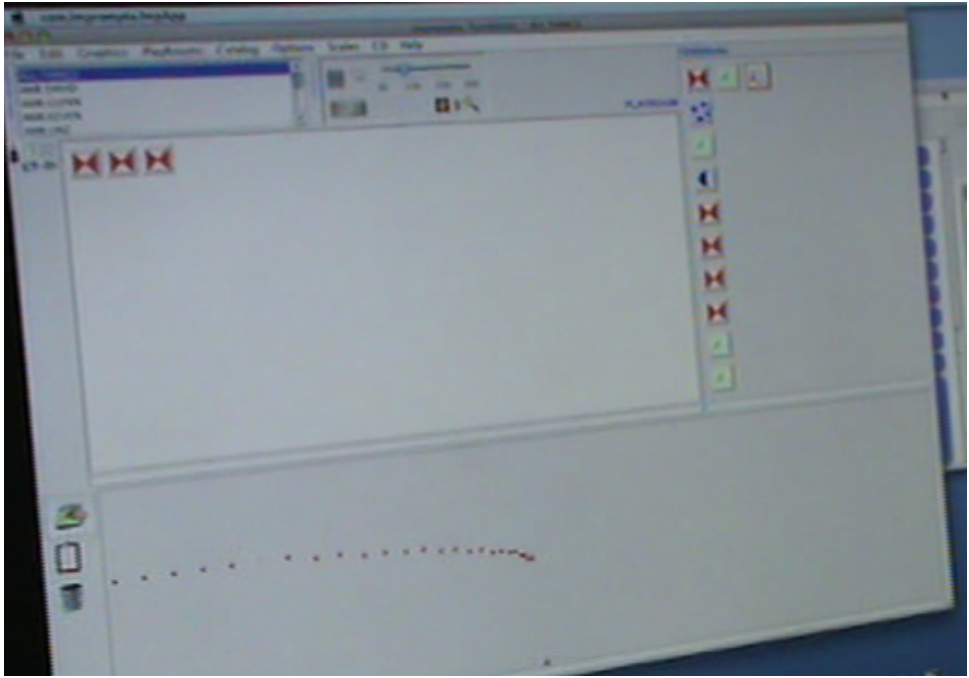

*Figure 2: Ariana's TuneBlocks Model of Acceleration*

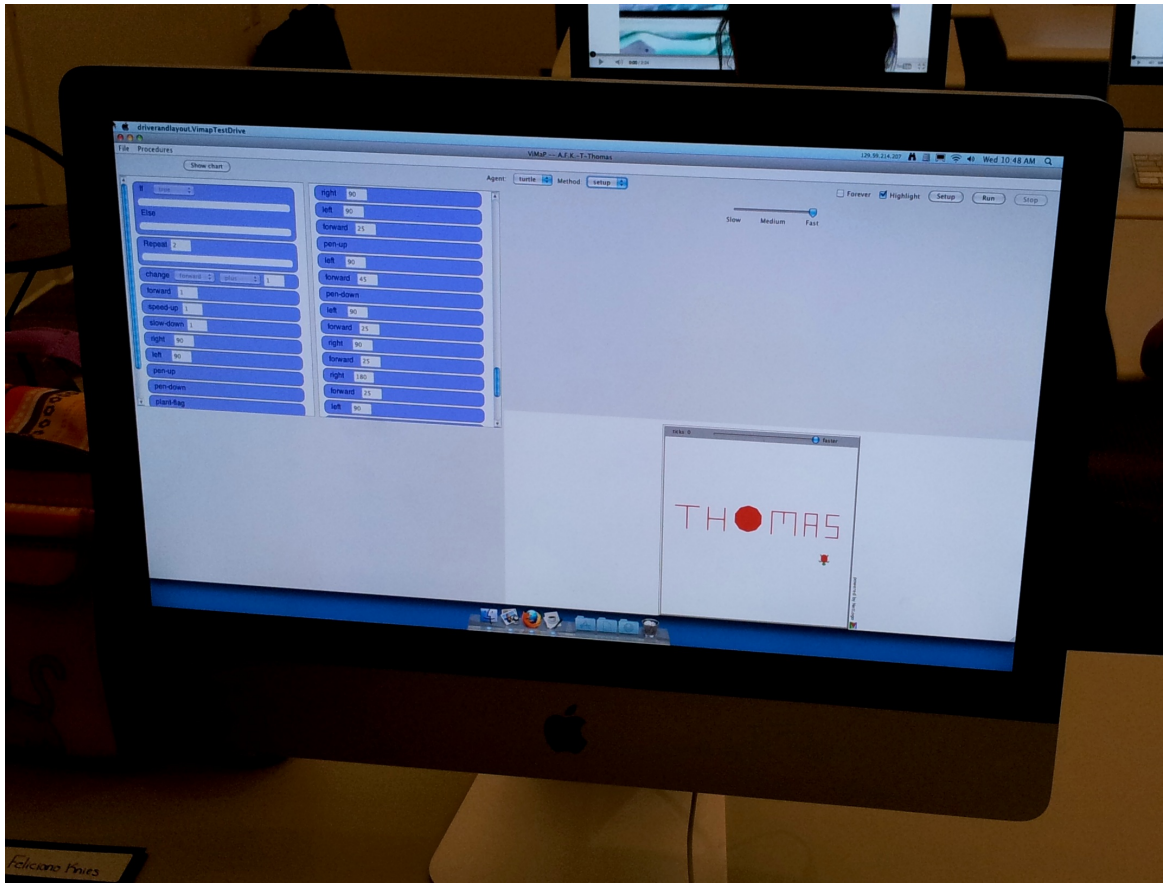

*Figure 3: Ariana's ViMAP model of "Thomas"*

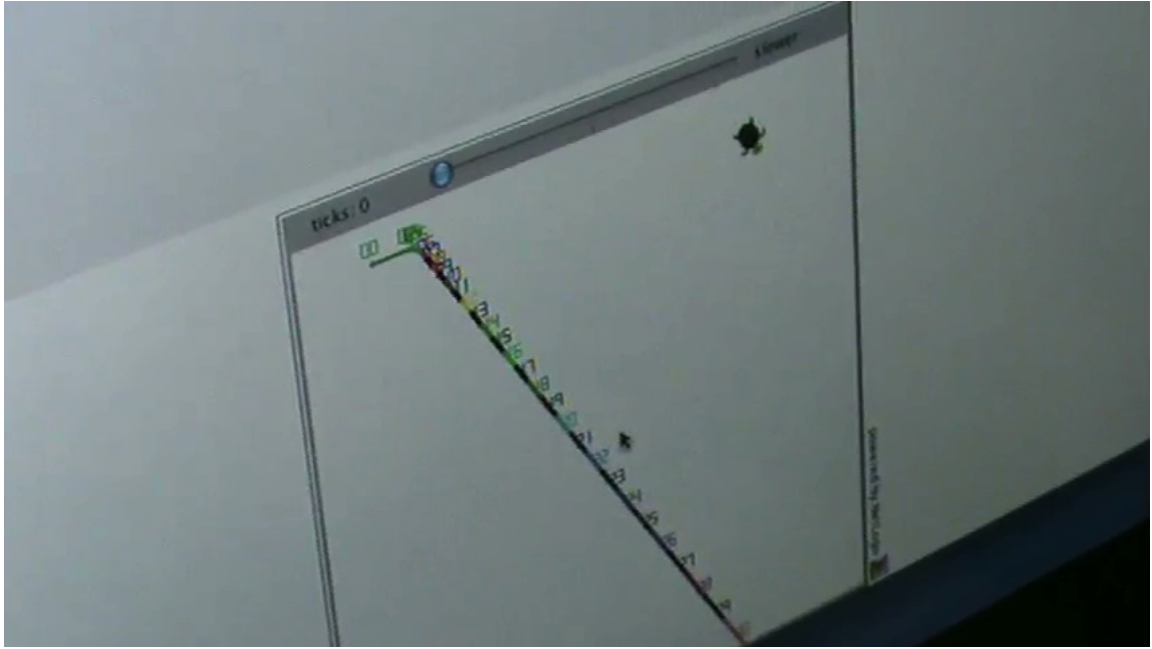

*Figure 4: Ariana and Matt's ViMAP model of motion*

[1] Andrea A. diSessa, *Changing Minds: Computers, Learning, and Literacy*. (MIT Press, 2001), 4.

[2] For examples, see Uri Wilensky, Corey E. Brady, and Michael S. Horn, "Fostering Computational Literacy in Science Classrooms," *Communications of the ACM* 57, no. 8 (August 2014): 24–28; Sengupta, Pratim, John S. Kinnebrew, Gautam Biswas, and Douglas Clark. "Integrating Computational Thinking with K-12 Science Education: A Theoretical Framework," *Education and Information Technologies* 18, no. 2 (2013): 351.

[3] Michael Eisenberg, "Technology for Learning: Moving from the Cognitive to the Anthropological Stance" (Paper in *Proceedings of the 10th International Conference of the Learning Sciences*, Sydney, 2012): 379.

[4] John Dewey, *Democracy and Education* (1916) *MW* 9, 82. The work *Democracy and Education* will be cited in the text as *DE* for all subsequent references.

[5] Chris Higgins, "Instrumentalism and the Clichés of Aesthetic Education: A Deweyan Corrective," *Education and Culture* 24, no. 1 (2008): 6-19.

[6] Peter Kosso, "The Omniscienter: Beauty and Scientific Understanding," *International Studies in the Philosophy of Science* 16, no. 1 (2002): 41–42.

[7] Gideon Engler, "Einstein and the Most Beautiful Theories in Physics," *International Studies in the Philosophy of Science* 16, no. 1 (2002): 27.

[8] S. Chandrasekhar. Truth and Beauty: Aesthetics and Motivations in Science. (Chicago: The University of Chicago Press, 1987): 64-73.

[9] Alexander Reuger, "Aesthetic appreciation of experiments: the case of 18th-centruy mimetic experiements," *International Studies in the Philosophy of Science* 16, no. 1 (2002): 51 - 53.

[10] David Gooding, "Varying the Cognitive Span: Experimentation, Visualization, and Computation" In H. Radder (Ed.), *The Philosophy of Scientific Experimentation* (Pittsburgh, PA, University of Pittsburgh Press, 2003): 261.

[11] Lorraine Daston and Peter Galison, *Objectivity* (Cambridge: MIT Press, 2010).

[12] Lorraine Daston and Peter Galison, "The image of objectivity," *Representations*, No. 40, Special Issue: Seeing Science (Autumn, 1992), pp. 81-128

[13] Mark Mcleod and Nancy Nerssessian, "Building Simulations from the Ground Up: Modeling and Theory in Systems Biology", Philosophy of Science, 80 (2013)

[14] Mark Girod, C. Rau, & A. Schepige. "Appreciating the Beauty of Science Ideas: Teaching for Aesthetic Understanding," *Science Education* 87 (2003): 574 - 587.

[15] Britt Jakobson and Per-Olof Wickman, "The Roles of Aesthetic Experience in Elementary School Science," *Research in Science Education* 38, no. 1 (2008): 45 - 66.

[16] Jay Lemke. "Articulating Communities: Sociocultural Perspectives on Science Education." *Journal of Research on Science Teaching* 38, no. 3 (2001) 300-301.

[17] Pickering, Andrew. *The Mangle of Practice: Time, agency, and science* (University of Chicago Press, 1995).

[18] Elinor Ochs, Patrick Gonzales, and Sally Jacoby. "When I come down I'm in the domain state: grammar and graphic representation in the interpretive activity of physicists." *Studies in Interactional Sociolinguistics* 13 (1996): 347-359; Evelyn Fox Keller. *A Feeling for the Organism: The Life and Work of Barb McClintock* (New York: W. H. Freeman and Company: 1983): 117-118.

[19] Lisa Delpit. "The Silenced Dialogue: Power and Pedagogy in Educating Other People's Children," Harvard Educational Review 58, no. 3 (August 1988): 280-283

[37] Transcript conventions include the following: [ ] Brackets are used to show overlapping speech of two speakers; = Latched speech; .. Pause, less than 2 seconds